\documentclass[prl,twocolumn,epsf,psfig]{revtex4}
\usepackage{graphicx}

\begin{document}

\title{Population imbalanced Fermi gases in quasi two dimensions}

\author{Theja N. De Silva}
\affiliation{Department of Physics, Applied Physics and Astronomy,
The State University of New York at Binghamton, Binghamton, New York
13902, USA.}
\begin{abstract}
We study s-wave pairing of population imbalanced Fermi atoms in quasi two dimensions using a mean field theory. At zero temperature, we map out the phase diagram in the entire Bardeen, Cooper
and Schrieffer-Bose Einstein condensation (BCS-BEC) crossover region by investigating the effect of weak atom tunneling between
layers. We find that the superfluid phase stabilizes as one decreases the atom tunneling between layers. This allows one to
control the superfluid-normal first order phase transition by tuning a single experimental parameter.
Further, we find that a tunneling induced polarized superfluid phase appears in a narrow parameter region in the BEC regime. At Finite temperatures, we use a
Landau-Ginzberg functional approach to investigate the possibility of spatially inhomogeneous Fulde-Ferrel-Larkin-Ovchinnikov (FFLO) phase in the weakly interacting BCS limit
near the tricritical point of spatially homogenous superfluid, FFLO, and normal phases. We find that the normal-FFLO phase transition is first order transition as opposed to
the continues transition predicted in zero temperature theories.
\end{abstract}

\maketitle

\section{I. Introduction}

Since its realization in dilute atomic gases, superfluidity of alkali atoms has been studied extensively in various externally controllable environments~\cite{ex}.
For the case of fermionic systems, superfluidity arises due to the Bose condensation of pairs of fermions at low temperatures. One of the most fascinating control parameters has been the
two-body scattering length between two atoms in two hyperfine spin states.
The two-body scattering length in three dimensions (3D) can be controlled dramatically by the use of magnetically tuned Feshbach resonance~\cite{fb}.
For dilute atomic systems at low temperatures, the two-body interaction is linearly proportional to the scattering length in 3D.
As a result of this proportionality, by controlling the two-body scattering length, the smooth crossover between the Bose-Einstein condensation (BEC) of strongly bound diatomic
molecules to the BCS limit of weakly bound cooper pairs has been observed experimentally. For a negative scattering length, the interaction between two atoms in two hyperfine spin states is attractive and
momentum space paring gives BCS superfluidity at low temperatures. For a positive scattering length, the interaction is repulsive and two body bound states exist in the vacuum
gives composite bosonic nature for two atoms paired in coordinate space. Bose Einstein condensation of these composite bosons gives superfluidity on the
BEC side of the resonance. These two regimes smoothly connect at unitarity where the scattering length is infinite in 3D. For attractive two body potentials in two dimensions (2D),
there always exists a two body bound state. The 2D bound state energy depends on both 3D scattering length and the laser intensity which used to create one dimensional
lattice to accommodate 2D layers. Therefore, 2D paring interactions can be controlled by tuning either the 3D scattering length or the laser intensity.

The most recent experiments with ultra-cold
Fermi gasses have been the focuss of
population imbalance which leads to the competition between superfluidity and magnetism~\cite{imr, immit}. Because of the
large spin relaxation time of the atoms, experimentalists were able to maintain a fixed
polarization $P=(N_{\uparrow}-N_{\downarrow})/(N_{\uparrow}+N_{\downarrow})$ over the entire time of the
experiments, here $N_{\uparrow/\downarrow}$ is the number of atoms in up/down pseudo spin state.

For the systems of atoms trapped in external harmonic potentials in three-dimensions, phase separation between normal phases and various superfluid phases has been
experimentally observed~\cite{imr, immit}. Theoretical investigation of population imbalanced fermion paring in three-dimensions (3D)
has been encouraged by these series of recent experiments~\cite{theory}. For spatially homogenous systems, various exotic phases, such as Sarma phase,
Fulde-Ferrel-Larkin-Ovchinnikov (FFLO) phase, a phase with deformed Fermi surfaces, and phase separation have been suggested~\cite{exotic}.
In trapped systems, various phases are separated into concentric shells and the shell structure depends on both the interaction strength and the polarization~\cite{imt1, imt2}.
The boundary between normal and superfluid regions depends on the trap geometry. Experiments done in high aspect ratio traps show that the superfluid-normal
boundary does not follow the equipotential contours of the trap and show significant distortion of the central superfluid region~\cite{imr}.
Quit remarkably, this distortion of the central superfluid shell can be explained by the surface tension between superfluid and normal regions~\cite{thprl, stoof, natu, thn}.
Local microscopic physics on the superfluid-normal boundary (due to the energy cost) causes this surface tension. The most recent theoretical
studies in 3D reveal the importance of the interaction in the normal phase to correctly explain
the experimentally observed Chandrasekhar-Clogston limit of critical polarization~\cite{recati, tezuka}.

The recent theoretical efforts of understanding the fermion pairing in two-dimensional population imbalanced systems
attempt to explore the phase diagram in the BCS-BEC crossover region~\cite{tempere, conduit, he}.
Experimentally, such a system can be created by applying a
relatively strong one-dimensional (1D) optical potential to an
ordinary three-dimensional system. Previous theoretical studies have
concentrated on the nature of various phases in 2D layers where the tunneling between layers are
neglected.

As the experimental setup in 2D is created out of an ordinary 3D system by applying a 1D lattice,
atom tunneling between layers are always present. This inter-layer
atomic tunneling can be controlled by a single parameter, namely the intensity of the 1D optical lattice. The purpose of the present
work is; (1) understand the effect of tunneling, (2) find out how one can control the phase transitions by using this easy controllability, and (3) study
possible FFLO phase in 2D. The motivation of studying FFLO phase came from the fact that FFLO phase is more favorable in low dimensions~\cite{lowDFFLO}. In 3D,
FFLO phase has not been seen in experiments, though theory predicts that such a phase can exist in a narrow parameter region in the weakly interacting limit.

The first part of this work is a natural
generalization of the work presented in refs.~\cite{tempere, conduit, he} to include the weak atom tunneling between atomic layers.
We find that the first order
superfluid-normal phase transition can be effectively controlled by the laser intensity. In typical
3D population imbalanced gases, one has to change the population imbalance to control the superfluid-normal phase transition.
Moreover, we find that the inclusion of tunneling allows to stabilize the polarized superfluid phase in BEC regime.
Further, in the weak coupling BCS limit, we investigate the possible
inhomogeneous FFLO phase near the tricritical point of superfluid, FFLO, and normal phases and find that the normal-FFLO phase transition is of first order.

The paper is organized as follows. In section II, we consider a zero temperature mean field theory and derive analytical expressions for the energy, density and gap
equation for weakly coupled layers. Then we predict the phase diagram in the entire BCS-BEC crossover region and discuss the effect of weak atom tunneling on
the superfluid-normal boundary. In section III, we neglect the coupling between layers and use a Landau-Ginzberg functional approach at finite temperatures. In
the weak coupling BCS limit, we derive an analytical expression for the Landau's free energy functional and discuss the possible FFLO phase near the
tricritical point. Finally, our conclusion is given in section IV.

\section{II. Zero temperature phase diagram}

We consider an interacting two-component Fermi atomic gas trapped in
quasi two dimensions. The 1D optical potential created
by the counter propagating laser beams has the form $V = sE_R\sin^2
(2\pi z/\lambda)$. Here $\lambda$ is the wavelength of the
laser beam, $E_R = \hbar^2(2\pi/\lambda)^2/(2M)$ is the recoil
energy. The dimensionless parameter $s$ can be used to modulate the
laser intensity. When the parameter $s$ is large, the atomic system
forms a stack of weakly coupled 2D planes with periodicity $d =
\lambda/2$. The Hamiltonian of the system $H = \sum_jH_j$ is
represented by,

\begin{eqnarray}\label{model1}
H_j = \int
d^2\vec{r}\biggr\{\sum_{\sigma}\psi^{\dagger}_{j\sigma}(r)[-\frac{\hbar^2\nabla^2_{2D}}{2M}-\mu_{\sigma}]\psi_{j\sigma}(r)
\\ \nonumber + t
\sum_{\sigma}[\psi^{\dagger}_{j\sigma}(r)\psi_{j+1\sigma}(r)+hc] \\
\nonumber
+U_{2D}\psi^{\dagger}_{j\uparrow}(r)\psi^{\dagger}_{j\downarrow}(r)\psi_{j\downarrow}(r)\psi_{j\uparrow}(r)\biggr\}
\end{eqnarray}

\noindent where $j$ is the layer index with $r^2 = x^2+y^2$,
$\nabla_{2D}$ is the 2D gradient operator and $U_{2D}$ is the 2D
interaction strength. The operator $\psi^{\dagger}_{j\sigma}(r)$
creates a fermion of mass $M$ in $j$th plane with pseudo spin $\sigma =
\uparrow, \downarrow$ at position $r = (x, y)$. We consider a tight
1D lattice where the atomic wave function becomes more and more
localized in the planes. Using the harmonic approximation around the
minima of the optical lattice potential~\cite{jaksch}, we find the
lattice tunneling parameter $t/E_R =
(2s^{3/4}/\sqrt{\pi})\exp[-2\sqrt{s}]$. Notice that the inter layer
tunneling energy $t$ can be varied by changing the laser intensity
parameter $s$. In the limit $t \rightarrow 0$, the system is
decoupled planes of Fermi atoms.

In this section, we consider zero temperature and use a mean filed
theory to decouple the interaction term writing
$U_{2D}\psi^{\dagger}_{j\uparrow}(r)\psi^{\dagger}_{j\downarrow}(r)\psi_{j\downarrow}(r)\psi_{j\uparrow}(r)
= [\Delta^{\dagger}
\psi_{j\downarrow}(r)\psi_{j\uparrow}(r)+h.c]-|\Delta|^2/U_{2D}$.
Further we neglect the possible inhomogeneous
Fulde-Ferrell-Larkin-Ovchinnikov paring and leave the finite
temperature FFLO discussion for the next section. The Fourier
transform of the Hamiltonian gives,

\begin{eqnarray}\label{model2}
H= \sum_{k_{\|},\sigma, m}(\epsilon_{k_{\|}} - \mu_{\sigma})a^{\dagger}_{m\sigma
}(k_{\|}) a_{m\sigma}(k_{\|}) \\ \nonumber+ t \sum_{k_{\|},\sigma,
m}[a^{\dagger}_{m+1\sigma}(k_{\|}) a_{m\sigma}(k_{\|}) + h.c] \\ \nonumber
+\sum_{k_{\|},m}[\Delta a^{\dagger}_{m\uparrow}(k_{\|})
a^{\dagger}_{m\downarrow}(-k_{\|})+h.c]-\frac{|\Delta|^2}{U_{2D}}
\end{eqnarray}

\noindent where $\epsilon_{k_{\|}} = \hbar^2k_{\|}^2/(2M)$ with $k_{\|}^2 =
k_x^2+k_y^2$. The notations $m$ and $k_{\|}$ represent the index of the plane and the momentum parallel to the 2D layers.
The operator $a^{\dagger}_{m\sigma}(k_{\|})$
creates a fermion of mass $M$ in $m$th plane with pseudo spin $\sigma =
\uparrow, \downarrow$ and momentum $k_{\|}$. The periodicity along the z-direction allows us to
write the Fermi operators,

\begin{eqnarray}\label{ft}
a_{m\uparrow}(k_{\|}) = \sum_{k_z}\exp[ik_zmd]c_{\uparrow}(k)
\nonumber \\
a_{m\downarrow}(-k_{\|}) = \sum_{k_z}\exp[-ik_zmd]c_{\downarrow}(-k)
\end{eqnarray}

\begin{figure}
\includegraphics[width=\columnwidth]{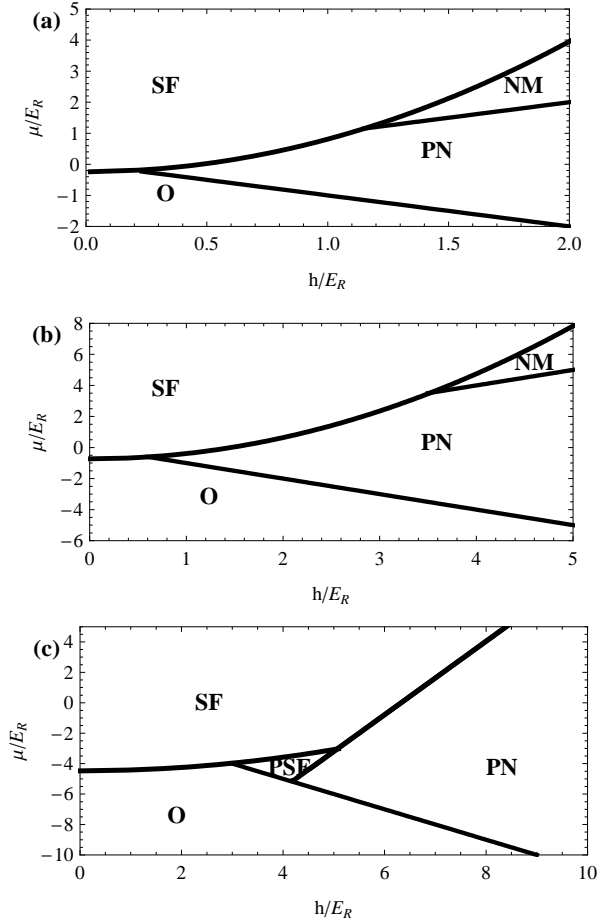}
\caption{Phase diagram of quasi 2D Fermi system in (a) BCS regime, (b) Unitarity regime,
and (c) BEC regime, where we fixed the 3D scattering length to
be $a_s = -\sqrt{\hbar^2/(2mE_R)}$, $a_s \rightarrow \infty$, and
$a_s = +\sqrt{\hbar^2/(2mE_R)}$ respectively. The lattice height is fixed by
setting $s=8$. The abbreviations are; SF: superfluid phase, PSF: polarized superfluid phase, NM:
normal mixed phase, PN: fully polarized normal phase, and O: vacuum
state.} \label{pd}
\end{figure}

\noindent where the fermi operator $c_{\sigma}^{\dagger}(-k)$ creates a fermion of momentum $k = (k_x, k_y, k_z)$ with pseudo spin $\sigma$.
Transforming the Hamiltonian given in Eq. (\ref{model2}) using above
canonical transformation followed by the usual Bogoliubov transformation, the
Hamiltonian per plane can be expressed in terms of quasi particle Fermi operators $\alpha^{\dagger}_{k\sigma}$ as

\begin{eqnarray}\label{ft}
H/N =
\sum_{k_{\|},k_z}\left(
                     \begin{array}{cc}
                       \alpha^{\dagger}_{k\uparrow} & \alpha_{k\downarrow} \\
                     \end{array}
                   \right)
\left(
                     \begin{array}{cc}
                       E_{k+} & 0 \\
                       0 & E_{k-} \\
                     \end{array}
                   \right)\left(
                            \begin{array}{c}
                              \alpha_{k\uparrow} \\
                              \alpha^{\dagger}_{k\downarrow} \\
                            \end{array}
                          \right)
\\
\nonumber+\sum_{k_{\|},k_z}[\bar{\epsilon}_k-\mu_{\downarrow}]-\frac{\Delta^2}{U_{2D}}
\end{eqnarray}

\noindent where $N$ is the number of planes, $\bar{\epsilon}_k =
\epsilon_{k_{\|}} + 2t \cos (k_zd)$, and $E_{k\pm} = -h \pm
\sqrt{(\bar{\epsilon}_k-\mu)^2+\Delta^2}$. Average chemical
potential $\mu = (\mu_{\uparrow}+\mu_{\downarrow})/2$ and chemical
potential difference $h = (\mu_{\uparrow}-\mu_{\downarrow})/2$. Without loss of generality, we take $h > 0$. The energy density per
plane $\Omega = (-1/\beta_0) \ln[Z_G]$ with $Z_G = Tr \exp[-\beta_0
H/N]$ is

\begin{eqnarray}\label{grand pot1}
\Omega = \sum_{k_z}\int \frac{d^2k_{\|}}{(2
\pi)^2}\biggr[\bar{\epsilon}_k-\mu-\sqrt{(\bar{\epsilon}_k-\mu)^2+\Delta^2}\biggr]\\
\nonumber -\frac{\Delta^2}{U_{2D}}
 - \frac{1}{\beta_0}\sum_{k_z}\int \frac{d^2k_{\|}}{(2
\pi)^2} \biggr[\ln[1 + \exp(-\beta_0 E_{k+})] \\ \nonumber + \ln[1 +
\exp(\beta_0 E_{k-})]\biggr]
\end{eqnarray}

\noindent Here $\beta_0 = 1/(k_BT)$ is the inverse temperature. Due to the nature of the contact interaction we used, Eq.~(\ref{grand pot1}) is
divergent so that proper regularization must be done. This can be done by relating the
2D contact interaction $U_{2D}$ to the bound state energy
$E_B$ as~\cite{u2d}

\begin{eqnarray}\label{u2d}
\frac{1}{U_{2D}} = - \int \frac{d^2k}{(2\pi)^2}
\frac{1}{\hbar^2k^2/M+E_B}.
\end{eqnarray}

\begin{figure}
\includegraphics[width=\columnwidth]{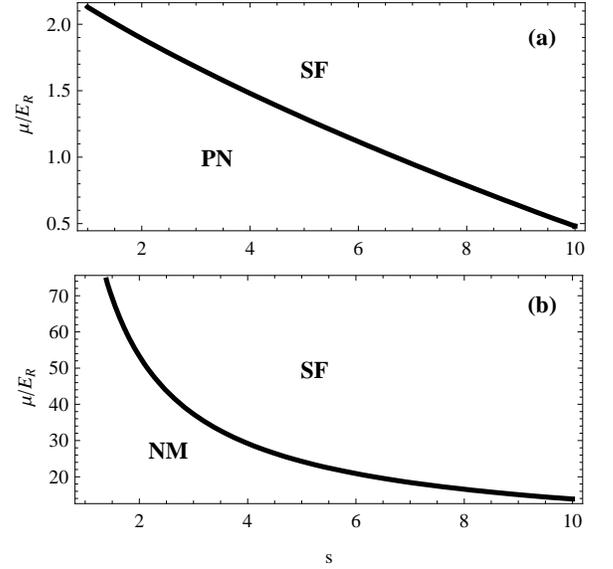}
\caption{Chemical potential on the superfluid phase and normal phase boundaries as a function of lattice height $s$. We fixed the
parameters in the BCS regime $a_s = -\sqrt{\hbar^2/(2mE_R)}$.
Panel (a) shows the chemical potential on the boundary between superfluid phase and the fully polarized normal phase at $h = E_R$ and
panel (b) shows the chemical potential on the boundary between superfluid phase and the normal mixed phase at $h = 4 E_R$. We find same qualitative behavior in the entire
BCS-BEC regime.} \label{boundary}
\end{figure}

\noindent The bound state energy is given by $E_B = (C\hbar \omega_L/\pi)
\exp[\sqrt{2\pi}l_L/a_s]$, where $a_s$ is the 3D s-wave scattering
length, $\omega_L=\sqrt{8 \pi^2 sE_R/(m \lambda^2)}$ is the trapping frequency due to the lattice, $l_L=\sqrt{\hbar/(m \omega_L)}$
is the oscillator length and $C \approx 0.915$~\cite{eb}. Notice that the bound state energy
depends not only on the 3D scattering length but also on the lattice
potential. After replacing 2D contact interaction $U_{2D}$ in Eq. (\ref{grand pot1}) by Eq.~\ref{u2d}, and then combining the first
and second term in Eq.~(\ref{grand pot1}), we remove the ultraviolet divergency. Converting the sum over $k_z$ into an integral over the first
brillouin zone and then performing the both momentum integrals, the energy density at
zero temperature is

\begin{eqnarray}\label{grand pot2}
\Omega =
\frac{m}{2\pi\hbar^2}\biggr\{\biggr(-\frac{\mu^2}{2}-\frac{\Delta^2}{4}-\frac{\mu}{2}\sqrt{\mu^2+\Delta^2}\\
\nonumber-\Theta(h-\Delta)h\sqrt{h^2-\Delta^2}\biggr)\\
\nonumber -\biggr(1 +
\frac{3\Delta^2\mu+2\mu^3}{2(\mu^2+\Delta^2)^{3/2}}\biggr)t^2
\\ \nonumber + \frac{15 \Delta^4 \mu}{8 (\mu^2+\Delta^2)^{7/2}}t^4
+ {\cal O}(t^6)\biggr\}
\end{eqnarray}

\noindent where Heaviside theta function $\Theta(x) = 1$ for $x>0$ and $0$ otherwise. Then the gap equation, the number density and density
difference are calculated by using the equations, $\partial
\Omega/\partial \Delta = 0$, $n = -\partial \Omega/\partial \mu$,
and $n_d = -\partial \Omega/\partial h$ respectively.

\begin{eqnarray}\label{gap}
\ln
\biggr[\frac{E_B}{-\mu+\sqrt{\mu^2+\Delta^2}}\biggr]-\Theta(h-\Delta)\ln
\biggr[\frac{h+\sqrt{h^2-\Delta^2}}{-h-\sqrt{h^2-\Delta^2}}\biggr]\\
\nonumber -\frac{\mu }{(\mu^2+\Delta^2)^{3/2}}t^2
+\frac{9\Delta^2\mu-6\mu^3}{4(\mu^2+\Delta^2)^{7/2}}t^4+ {\cal
O}(t^6)=0
\end{eqnarray}

\begin{eqnarray}\label{density}
n =
\frac{m}{2\pi\hbar^2}\biggr\{\biggr(\mu+\sqrt{\mu^2+\Delta^2}\biggr)+
\frac{\Delta^2}{(\mu^2+\Delta^2)^{3/2}}t^2
\\ \nonumber -\frac{3\Delta^2(\Delta^2-4\mu^2)}{\mu^2+\Delta^2)^{7/2}}t^4
+ {\cal O}(t^6)\biggr\}
\end{eqnarray}

\begin{eqnarray}\label{density}
n_d = \frac{m}{\pi\hbar^2}\Theta(h-\Delta)\sqrt{h^2-\Delta^2}
\end{eqnarray}

\begin{figure}
\includegraphics[width=\columnwidth]{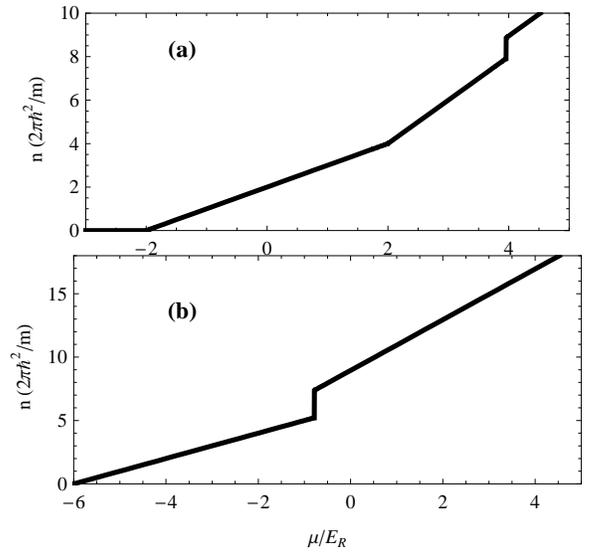}
\caption{The density of a quasi 2D Fermi system as a function of
chemical potential $\mu$ in (a) BCS regime [$a_s = -\sqrt{\hbar^2/(2mE_R)}$, and $h = 2E_R$] and (b) BEC regime
[$a_s = +\sqrt{\hbar^2/(2mE_R)}$, and $h = 6E_R$]. The lattice height is fixed by setting $s=8$.} \label{densityfig}
\end{figure}

\noindent We solve Eq.~(\ref{gap}) for given $s$, $a_s$, $\mu$ and $h$
for $\Delta$ and then determine the energy of the system from
Eq.~(\ref{grand pot2}). Comparing the energy of the system, we then map out
the phase diagram in $\mu-h$ parameter space as shown in
FIG.~\ref{pd}. As can be seen in FIG.~\ref{pd}, the qualitative feature of the phase diagrams in the
entire BCS-BEC regime is the same. However the normal mixed phase
(where both spin up and down components coexist) narrows down as one
goes from BCS regime to BEC regime. In trapped systems, h is uniform within the cloud, while chemical potential $\mu$
varies monotonically from the center to edge of the trap. Taking vertical slices of the phase diagram at fixed $h$,
one can see that the atomic cloud passes through various local phases giving a
shell structure in a trapped system. The shell structure depends on
both interaction strength $E_B$ and the chemical potential
difference $h$. The similar phase diagram in the $\mu-h$ plane is
obtained in the limit of $s \rightarrow \infty$ in Ref.~\cite{tempere}
and Ref.~\cite{conduit}. The main difference is the superfluid
phase region in the phase diagram narrows down as a result of weak tunneling (due to the finite $s$).
In contrast to 3D population imbalanced Fermi
systems~\cite{imt1}, the parameter region for polarized superfluid phase (both superfluid and normal coexisted
phase) in the BEC regime is very small. Using a very similar mean field analysis, Ref.~\cite{tempere}
and Ref.~\cite{conduit} conclude that the polarized superfluid phase
is absent in the limit of $s \rightarrow \infty$ in 2D. However, using a beyond mean field approach, Ref.~\cite{he}
predicts a polarized superfluid phase even in the $s \rightarrow \infty$ limit. Nevertheless, within our mean field description,
we find a tunneling induced polarized superfluid phase only in the BEC regime. Notice that we have constructed the phase diagram by solving the
equations for fixed $\mu$ and $h$. Alternatively, one can use the canonical ensemble (where $n$ and $n_d$ are held fixed)
to construct the phase diagram in $n$-$n_d$ plane. As $n_d$ is zero in the SF phase, a line in $n$-$n_d$ plane
will represent the SF phase. Between the SF line and PN/NM phase separates a region of phase separation~\cite{conduit}.

Figure~\ref{boundary} shows the variation of the chemical potential on the superfluid-normal boundaries. As the
lattice height increases, the superfluid phase stabilizes against
the normal phases. As a result, the superfluid cloud extends toward
the edge as one increases the 1D lattice height in a trapped system.
This is because, 2D interaction strength increases with increasing
the 1D lattice height. As the lattice height is controllable though the laser intensity, the laser intensity
can be considered as a non-destructive experimental knob to control the first order superfluid-normal
phase transitions as well as polarized superfluid phase. This
easy controllability is available only in quasi 2D systems and one has to change the population
imbalance in 3D systems to control the transition.

In figure~\ref{densityfig}, we present typical density profiles for two
different representative values of interaction strengths in the BCS
regime and BEC regime. For qualitative understanding, the chemical
potential axis can be considered as a spatial coordinate in trapped
systems. This is because, as we mentioned before, the chemical potential varies monotonically from the center to edge of the
cloud. The density
profile in the BCS regime shows a kink and a discontinuity
representing two phase boundaries. The discontinuity represents the
phase boundary between superfluid phase and the mixed normal phase
while the kink represents the phase boundary between mixed normal
phase and the fully polarized normal phase. In contrast, the density
profile in the BEC regime shows a discontinuity showing a
two-shell structure at given $h$ and $s$.

\section{III. FFLO phase near tricritical point}

\begin{figure}
\includegraphics[width=\columnwidth]{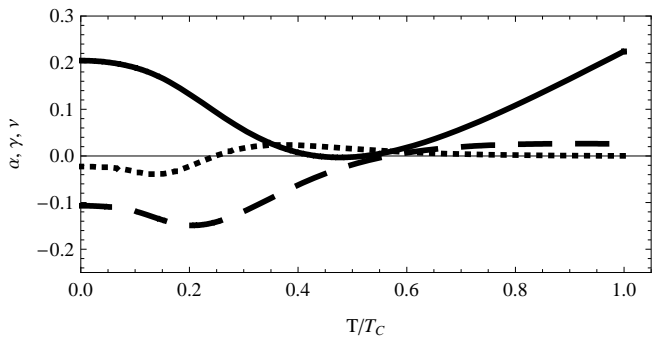}
\caption{The coefficients $\alpha$ (solid line), $\gamma$ (dashed line), and $\nu$ (dotted line) as a function of temperature.
We fixed the value of chemical potential difference $h = 1.08k_BT_C$ such that there are three phase transitions as lower the temperature.
See also FIG.~\ref{pdF}.} \label{para}
\end{figure}

In this section, we consider a finite temperature spin imbalanced Fermi system in two dimensions and
neglect atom tunneling between layers. The Hamiltonian of
the system in the mean field description reads,

\begin{eqnarray}\label{model2D}
H = \int
d^2\vec{r}\biggr\{\sum_{\sigma}\psi^{\dagger}_{\sigma}(r)[-\frac{\nabla^2_{2D}}{2M}-\mu]\psi_{\sigma}(r)
\\ \nonumber
+\sigma h\psi^{\dagger}_{\sigma}(r)\psi_{\sigma}(r)
+\Delta(r)\psi^{\dagger}_{\uparrow}(r)\psi^{\dagger}_{\downarrow}(r)+h.c\biggr\}
\end{eqnarray}

\noindent where $\sigma$ is the pseudo spin $\uparrow(+)$ and $\downarrow(-)$.
In order to study the FFLO state near the tricritical point, we use a
Landau's phenomenological approach to write the free energy
functional in the weakly interacting BCS limit. As shown in Ref.~\cite{u2d}, the chemical potential
of a 2D Fermi gas is given by $\mu =\epsilon_F-E_B/2$, where $\epsilon_F$ and $E_B$ are Fermi energy
and binding energy of the two body bound state. In the weak coupling BCS limit, $E_B \ll \epsilon_F$,
so that the chemical potential can be approximated by the Fermi energy. Further, spatial modulation of the order
parameter $\Delta (r)$ is small close to the tricritical point. Therefore, free energy
can be expanded in powers of the superfluid order parameter. The
coefficients of the terms in each order determine the nature of the
each phase transition. Following the reference~\cite{lg}, the
thermodynamic potential can be written as,

\begin{eqnarray}\label{LG}
F = \alpha |\Delta|^2 + \gamma v_F^2|\partial \Delta|^2/2 + \gamma
|\Delta|^4 \\ \nonumber + \nu \{|\Delta|^6 +
3v_F^4|\partial^2\Delta|^2/16 + 2v_F^2|\Delta|^2|\partial \Delta|^2
\\ \nonumber + v_F^2[(\Delta^\dagger)^2(\partial \Delta)^2+\Delta^2(\partial
\Delta^\dagger)^2]/4\}
\end{eqnarray}

\noindent with the coefficients $\alpha = N(0) \{\ln (T/T_c) + Re
\psi[1/2 + ih/(2\pi T)]-\psi(1/2)\}$, $\gamma=N(0)\pi K_3/4$ and
$\nu=-N(0)\pi K_5/8$. Here $\psi$ is the digamma function, $T$ is
the temperature, $T_c$ is the critical temperature, $N(0)$ is the
density of states at the Fermi surface, $v_F$ is the Fermi velocity,
and the functions $K_3$ and $K_5$ are given by,

\begin{eqnarray}\label{LG}
K_3 = -\frac{1}{8\pi^3T^2}Re[\psi^{(2)}(x)] \\ \nonumber K_5 =
-\frac{1}{384\pi^5T^4}Re[\psi^{(4)}(x)]
\end{eqnarray}
\noindent Here $x=1/2-ih/(2\pi k_BT)$ and $\psi^{(n)}(x)$ is the n-th derivative of digamma function on
its argument. Notice that the forth order term $\gamma$
simultaneously accompanied by the gradient term ($|\partial \Delta|^2$) in the BCS
mean field theory so that we must include the sixth order term
$\nu$. When both $\alpha$ and $\gamma$ are positive, thermodynamic
potential gives a single minimum at $\Delta=0$ and the system is in
the normal state. We consider the simplest FFLO state where the Cooper
pairs in the coordinate space have the plane wave form
$\Delta(r)=\Delta_0 \exp[i\vec{q} \cdot \vec{r}]$. When $\gamma$
becomes negative, the modulated order has the lowest energy. The
tricritical point is determined by the conditions $\alpha =0$ and
$\gamma=0$. The second order phase transition from normal state to
homogenous superfluid state is determined by the condition
$\alpha=0$. Above the superfluid transition ($T>T_C)$, $\alpha >0$
and below the transition $\alpha <0$ (see FIG.~\ref{para} and discussion below). For the first order homogenous
superfluid-normal phase transition, we find the value of $\Delta$ at
the transition by setting $\partial F/\partial\Delta = 0$ at $q=0$.
This leads to the solution
$\Delta_{\pm}=[-\gamma\pm\sqrt{\gamma^2-3\alpha \nu}]/(3\nu)$. The
condition for the first order transition is then determined by
setting $F(\Delta=0) = F(\Delta_+)$ which gives
$\gamma=-2\sqrt{\alpha \nu}$.

\begin{figure}
\includegraphics[width=\columnwidth]{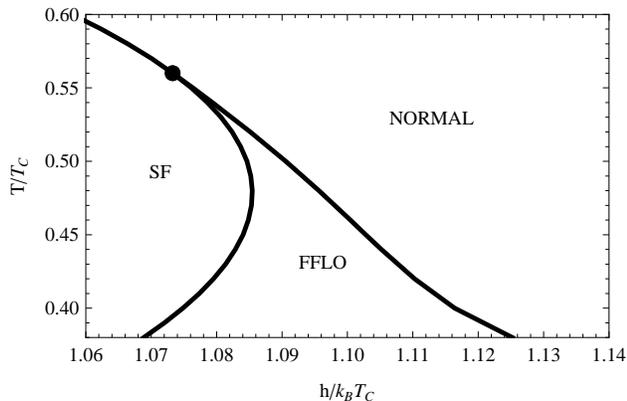}
\caption{Phase diagram of 2D population imbalance Fermi system near
the tricritical point. All three phases, superfluid (SF), normal,
and inhomogeneous FFLO coexist at the tricritical point represented by
the solid circle. } \label{pdF}
\end{figure}

Let us now consider the inhomogeneous superfluid-normal phase
transition with finite $q$. Using above simple anzats for the order parameter $\Delta$, the free
energy is given by $F = (\alpha + \gamma Q^2/2 + 3 \nu Q^4/16)\Delta^2 +
(\gamma + 3 \nu Q^2/2)\Delta^4+\nu \Delta^6$, where $Q=v_Fq$. For a
possible second order phase transition, we set the coefficient of
$\Delta^2$ to be zero, and then minimize it with respect to $Q$.
This gives the condition for possible second order transition into
FFLO state; $\gamma=-2\sqrt{3\alpha \nu/4}$ with center of mass
paring momentum $q=\sqrt{-8 \gamma/(6\nu)}/v_F$.

For a possible first order transition into FFLO state, we minimize
the free energy with respect to both $\Delta$ and $Q$. This gives
the condition $\gamma=-2\sqrt{6\alpha\nu/5}$ with paring momentum
$q=\sqrt{-2\gamma/(6\nu)}/v_F$.

Transforming these second order and
first order phase transition conditions into temperature ($T$) and
chemical potential difference ($h$) through $T-h$ dependence of the parameters,
$\alpha$, $\gamma$, and $\nu$, we find that the first order phase transition line is always appears
before the second order phase transition line in $T-h$ plane as one decreases $h$. This
indicates that the transition from normal phase into FFLO phase in a 2D
population imbalanced system at finite temperature is first order. This contrast to the zero temperature
theories~\cite{conduit,zerotFFLO}, where thermodynamic potential is
expanded only to the second order in $\Delta$. In these zero temperature
theories, the transition from normal state into FFLO state is found to be a
second order in 2D. Notice, here we used the simplest FFLO state which has the plane wave form. We do not expect
that the nature of the transition would change if we choose somewhat complex forms of the FFLO state.

\begin{figure}
\includegraphics[width=\columnwidth]{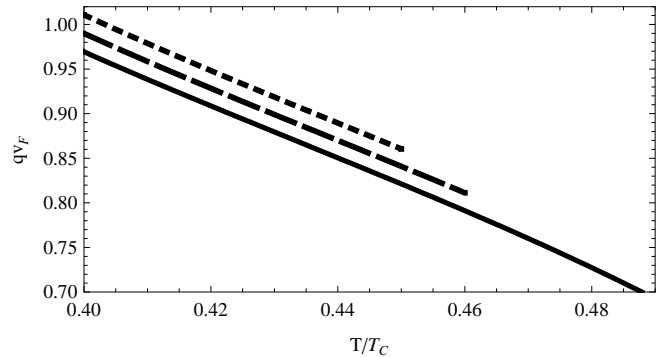}
\caption{Center of mass paring momentum as a function of temperature
for $h/(k_BT_C) = 1.11$ (dotted line), $h/(k_BT_C) = 1.10$ (dashed
line), and $h/(k_BT_C) = 1.09$ (solid line). } \label{qp}
\end{figure}

The phase diagram in the vicinity of tricritical point is shown in
FIG.~\ref{pdF}. As our theory is not valid at low temperatures, we
present our results at a finite temperature range. In FIG.~\ref{para}, we show the temperature dependence of
the parameters $\alpha$, $\gamma$, and $\nu$. As these parameters are functions of both temperature and chemical
potential difference $h$, we choose $h ( \equiv 1.08k_BT_C)$ such that there are three phase transitions as one lowers the temperature.
The sequence of these phase transitions are Normal $\rightarrow$ FFLO $\rightarrow$ SF $\rightarrow$ FFLO.
The transition temperatures for Normal-FFLO, FFLO-SF and SF-FFLO are 0.538$T_C$, 0.533$T_C$, and 0.434$T_C$ respectively.
Notice that the parameter $\alpha$ is negative in the SF phase and positive in both normal and FFLO phases.
Further, $\gamma$ changes its sign as the system undergos first order transition into FFLO state. The center of
mass pairing momentum in the FFLO phase is shown in FIG.~\ref{qp}
for three different representative values of chemical potential differences. As shown in the phase diagram in FIG.~\ref{pdF},
the FFLO window always appears in between superfluid and normal phases. Therefore, in trapped systems, the FFLO phase will
appear bordering between superfluid and normal concentric shells at low temperatures below tricritical point.

\section{IV. Conclusions}

We studied Fermi superfluidity in quasi two dimensions focussing on population imbalance and weak atom tunneling between layers.
By using a mean field theory, we derived an analytical expressions for the free energy, density, and solution of the gap equations
at zero temperature. Analyzing the free energy of the system, we mapped out the phase diagram where we find phase separation due
to the population imbalance. Further, we find that the suppression of weak atomic tunneling stabilizes the superfluid phase due to the enhancement of the 2D
interaction strength. The easy controllability of the tunneling through the laser intensity allows one to tune the first order superfluid-normal
phase transition. We also find that the polarized superfluid phase which is absent in pure 2D limit in the mean field description,
can be realized by allowing atoms to tunnel between layers in the BEC limit.
The tunneling dynamic of the system can be understood by generalizing the results in reference~\cite{th1D}.

At finite temperature, we used a Landau's functional
approach in the weak coupling BCS limit and discussed the inhomogeneous FFLO phase near the tricritical point. We find that
as one decreases the population imbalance, system undergoes a first order phase transition into FFLO phase where zero temperature theories predict a second
order phase transition.

\section{V. Acknowledgments}
This work was supported by the Binghamton University. We are very grateful to Michael Lawler for very
enlightening discussions and critical comments on the
manuscript.


\begin{references}

\bibitem{ex}M. H. Anderson, J. R.
Ensher, M. R. Matthews, C. E. Wieman, E. A. Cornell, Science \textbf{269}, 198, (1995); K.B. Davis, M.-O. Mewes, M.R. Andrews, N.J. van
Druten, D.S. Durfee, D.M. Kurn, and W. Ketterle, Phys. Rev. Lett. \textbf{75}, 3969, (1995); C. C. Bradley,
C. A. Sackett, J. J. Tollett, and R. G. Hulet, Phy. Rev. Lett. \textbf{75}, 1687 (1995); B. DeMarco, D. S. Jin, Science \textbf{285}, 1703
(1999); A. G. Truscott,
K. E. Strecker, W. I. McAlexander, G. B. Partridge, R. G. Hulet, Science
\textbf{291}, 2570 (2001); K.
M. O'Hara, S. L. Hemmer, M. E. Gehm, S. R. Granade, J. E. Thomas, Science \textbf{298}, 2179,
(2002); C. A. Regal, M. Greiner, and D. S. Jin, Phys. Rev. Lett. \textbf{92}, 040403 (2004); M. W. Zwierlein, C. A. Stan, C. H. Schunck, S. M. F. Raupach,
A. J. Kerman, and W. Ketterle, Phys. Rev. Lett. \textbf{92}, 120403 (2004); C. Chin et al., Science \textbf{305}, 1128 (2004); T. Bourdel, L. Khaykovich, J. Cubizolles, J. Zhang,
F. Chevy, M. Teichmann, L. Tarruell, S. J. J. M. F. Kokkelmans, and C. Salomon, Phys. Rev. Lett. \textbf{93}, 050401 (2004);
J. Kinast, S. L. Hemmer, M. E. Gehm, A. Turlapov, and J. E. Thomas, Phys. Rev. Lett. \textbf{92}, 150402 (2004);
J J. Kinast, S. L. Hemmer, M. E. Gehm, A. Turlapov, and J. E. Thomas, Phys. Rev. Lett. \textbf{92}, 150402 (2004);
M. Bartenstein, A. Altmeyer, S. Riedl, S. Jochim, C. Chin, J. H. Denschlag, R. Grimm, Phys. Rev. Lett. \textbf{92}, 203201 (2004).

\bibitem{fb} U. Fano, Phys. Rev. A \textbf{124}, 1866 (1961); H Feshbach,
Ann. Phys. \textbf{5}, 357 (1961).

\bibitem{imr} G. B. Partridge, W. Li, R. I. Kamar, Y. Liao, and R. G. Hulet, Science, \textbf{311}, 503 (2006);
G. B. Partridge, W. Li, R. I. Kamar, Y. Liao, and R. G. Hulet, Phys. Rev. Lett.  \textbf{97}, 190407 (2006).

\bibitem{immit} M. W. Zwierlein, A. Schirotzek, C. H. Schunck,
W. Ketterle, Science, \textbf{311}, 492 (2006); M. W. Zwierlein, A. Schirotzek, C. H. Schunck, W. Ketterle, Nature \textbf{442}, 54 (2006).

\bibitem{theory} For reviews see for examples; D. E. Sheehy,  and L. Radzihovsky, Ann. Phys, \textbf{322},1790 (2007); S. Giorgini, L. P. Pitaevskii, and S. Stringari,
Rev. Mod. Phys. \textbf{80}, 1215 (2008); and references there in.


\bibitem{exotic} G. Sarma, J.Phys.Chem.Solids \textbf{24}, 1029 (1963); I. Shovkovy and M. Huang, Phys. Lett. B \textbf{564}, 205 (2003);
M. Huang and I. Shovkovy, Phys.Rev. D \textbf{70}, 094030 (2004); W. Liu and F. Wilczek, Phys. Rev. Lett. \textbf{90}, 047002 (2003); P. Fulde and R. Ferrel, Phys. Rev. A \textbf{135}, 550 (1964);
A. Larkin and Y. Ovchinnikov, Sov. Phys. JETP \textbf{20},762 (1965); H. Muther and A. Sedrakian, Phys. Rev. Lett. \textbf{88}, 252503 (2002); A. Sedrakian et al:, Phys. Rev. A \textbf{72}, 013613 (2005);
P. Bedaque, H. Caldas and G. Rupak, Phys. Rev. Lett. \textbf{91}, 247002 (2003); H. Caldas, Phys. Rev. A \textbf{69}, 063602 (2004).


\bibitem{imt1} T. N. De Silva and E. J. Mueller, Phys. Rev. A \textbf{73}, 05602(R) (2006).

\bibitem{imt2} M. Haque, H. T. C. Stoof, Phys. Rev. A \textbf{74}, 011602 (2006).

\bibitem{thprl} T. N. De Silva and E. J. Mueller, Phys. Rev. Lett. \textbf{97}, 070402, (2006).

\bibitem{stoof} M. Haque and H. T. C. Stoof, Phy. Rev. Lett. \textbf{98}, 260406 (2007)

\bibitem{natu}  S. S. Natu and E. J. Mueller, arXiv:0802.2083 (2008).

\bibitem{thn}  S. K. Baur,  S. Basu,  T. N. De Silva, and E. J. Mueller, arXiv:0901.2945 (2009).

\bibitem{recati} A. Recati, C. Lobo and, S. Stringari, Phys. Rev. A \textbf{78}, 023633 (2008).

\bibitem{tezuka} M. Tezuka and M. Ueda, arXiv:0811.1650 (2008).

\bibitem{tempere} J. Tempere, M. Wouters, and J. T. Devreese, Phys. Rev. B \textbf{75}, 184526
(2007).

\bibitem{conduit} G. J. Conduit, P. H. Conlon, and B. D. Simons, Phys. Rev. A \textbf{77}, 053617
(2008).

\bibitem{he} L. He and P. Zhuang, Phys. Rev. A \textbf{78}, 033613 (2008).

\bibitem{lowDFFLO} M. M. Parish, S. K. Baur, E. J. Mueller, and D. A. Huse, Phys. Rev. Lett. \textbf{99}, 250403 (2007); H. Hu, X. Liu, and P. D. Drummond, Phys. Rev. Lett. \textbf{98}, 070403 (2007).

\bibitem{jaksch} D. Jaksch, C. Bruder, J. I. Cirac, C. W. Gardiner, and P.
Zoller, Phys. Rev. Lett. \textbf{81}, 3108 (1998).

\bibitem{u2d}M. Randeria, J. Duan, and L. Shieh, Phys. Rev. B \textbf{41}, 327
(1990); S. K. Adhikari, Am. J. Phys. \textbf{54}, 362 (1986).

\bibitem{eb} D. S. Petrov and G. V. Shlyapnikov, Phys. Rev. A \textbf{64}, 012706 (2001).

\bibitem{lg} A. I. Buzdin and H. Kachkachi, Physics letters A \textbf{225},
341 (1997).

\bibitem{th1D} T. N. De Silva, Phys. Rev. A \textbf{79}, 013612 (2009).

\bibitem{zerotFFLO} R. Combescot and C. Mora, Eur. Phys. J. B \textbf{44}, 189 (2005)



\end{references}
\end{document}